\documentclass{ceurart}

\usepackage{amsmath}
\usepackage{algorithm}
\usepackage{algorithmic}
\usepackage{comment}
\usepackage{listings}
\usepackage{graphicx}
\usepackage{subcaption}
\usepackage{booktabs}
\begin{document}

\copyrightyear{2026}
\copyrightclause{Copyright for this paper by its authors.
  Use permitted under Creative Commons License Attribution 4.0
  International (CC BY 4.0).}

\conference{Published in the Proceedings of the Workshops of the EDBT/ICDT 2026 Joint Conference (March 24-27, 2026), Tampere, Finland}

\title{Navigating the Shift: A Comparative Analysis of Web Search and Generative AI Response Generation}

\author[1]{Mahe Chen}[email=mahe.chen@mail.utoronto.ca]
\cormark[1]

\author[1]{Xiaoxuan Wang}[email=xxuan.wang@mail.utoronto.ca]

\author[1]{Kaiwen Chen}[email=kckevin.chen@mail.utoronto.ca]

\author[1]{Nick Koudas}[email=koudas@cs.toronto.edu]

\address[1]{Department of Computer Science, University of Toronto, Canada}

\cortext[1]{Corresponding author.}

\begin{abstract}
The rise of generative AI as a primary information source presents a paradigm shift from traditional web search. This paper presents a large-scale empirical study quantifying the fundamental differences between the results returned by Google Search and leading generative AI services. We analyze multiple dimensions, demonstrating that AI-generated answers and web search results diverge significantly in their consulted source domains, the typology of these domains (e.g., earned media vs. owned, social), query intent, and the freshness of the information provided. We then investigate the role of LLM pre-training as a key factor shaping these differences, analyzing how this intrinsic knowledge base interacts with and influences real-time web search when enabled. Our findings reveal the distinct mechanics of these two information ecosystems, leading to critical observations on the emergent field of Answer Engine Optimization (AEO) and its contrast with traditional Search Engine Optimization (SEO).
\end{abstract}


\maketitle

\section{Introduction}

The paradigm of information retrieval is undergoing its most significant transformation in decades. For years, users seeking information online have turned to search engines like Google, which act as gateways to a curated list of web documents. This process, governed by the principles of Search Engine Optimization (SEO), requires users to actively scan, evaluate, and synthesize information from multiple sources. The recent advent of generative AI-powered answer engines promises a radical alternative: delivering direct, synthesized answers to user queries.

While this shift offers unprecedented convenience, it also raises critical questions about the underlying mechanics and biases of this new information ecosystem. How do the answers provided by generative AI services fundamentally differ from the results of a traditional web search? The answer lies not just in the presentation, but in the provenance, typology, and recency of the sources they consult, as well as the intrinsic knowledge acquired during their Large Language Model (LLM) pre-training phase. Pre-training on vast, static web corpora creates a latent knowledge base that may prioritize established information over fresh content and influence which external sources are deemed necessary to cite for a given query.

In this paper, we present a large-scale empirical study to dissect these differences. We systematically compare the outputs of GPT-4o, Claude 4.5 Sonnet, and Perplexity Sonar Pro against Google Search and Gemini 2.5 Flash across two critical axes: querying for popular entities with abundant pre-training data and niche entities with limited data. Our analysis focuses on (1) the domains consulted and their classification; (2) the freshness of the cited URLs; and (3) the demonstrable effect of LLM pre-training on the final answer. Through this multi-faceted comparison, we reveal the distinct logics of ``web search'' and ``AI search.'' We conclude by contrasting the emergent strategies for Answer/Generative Engine Optimization (AEO/GEO) \cite{aggarwal2024geogenerativeengineoptimization, wan2024evidencelanguagemodelsconvincing, kumar2024manipulatinglargelanguagemodels,chen2025generativeengineoptimizationdominate} with traditional SEO, outlining the new landscape of information discoverability. Our findings provide a crucial foundation for understanding the reliability, biases, and economic implications of generative AI as an information source.

\section{Comparative Analysis of Search Results}

\subsection{Diversity and Overlap of Results}

\noindent \textit{Domain overlap.}
To quantify how AI-powered answer engines diverge from traditional web search, we begin by measuring the domain-level overlap between their cited sources and those returned by Google. This overlap provides a first-order indicator of whether the two paradigms draw from a shared information ecosystem or surface distinct domain spaces. We evaluate 1,000 ranking-style queries (e.g., ``Top 10 most reliable smartphones,'' ``Best reviewed airlines this season''), an important and growing commercial use case for AI search \cite{chatterji2025howpeopleusechatgpt} spanning ten consumer topics\footnote{smartphones, athletic shoes, skin care, electric cars, streaming services, laptops, airlines, hotels, credit cards, and smartwatches.} across five systems: Google Search, GPT-4o (web-enabled), Claude 4.5 Sonnet (web-enabled), Gemini 2.5 Flash (with Google Search grounding), and Perplexity Sonar Pro (search mode: web).

\begin{figure}[t]
    \centering
    \begin{subfigure}[t]{0.49\linewidth}
        \centering
        \includegraphics[width=\linewidth,height=4.2cm,keepaspectratio]{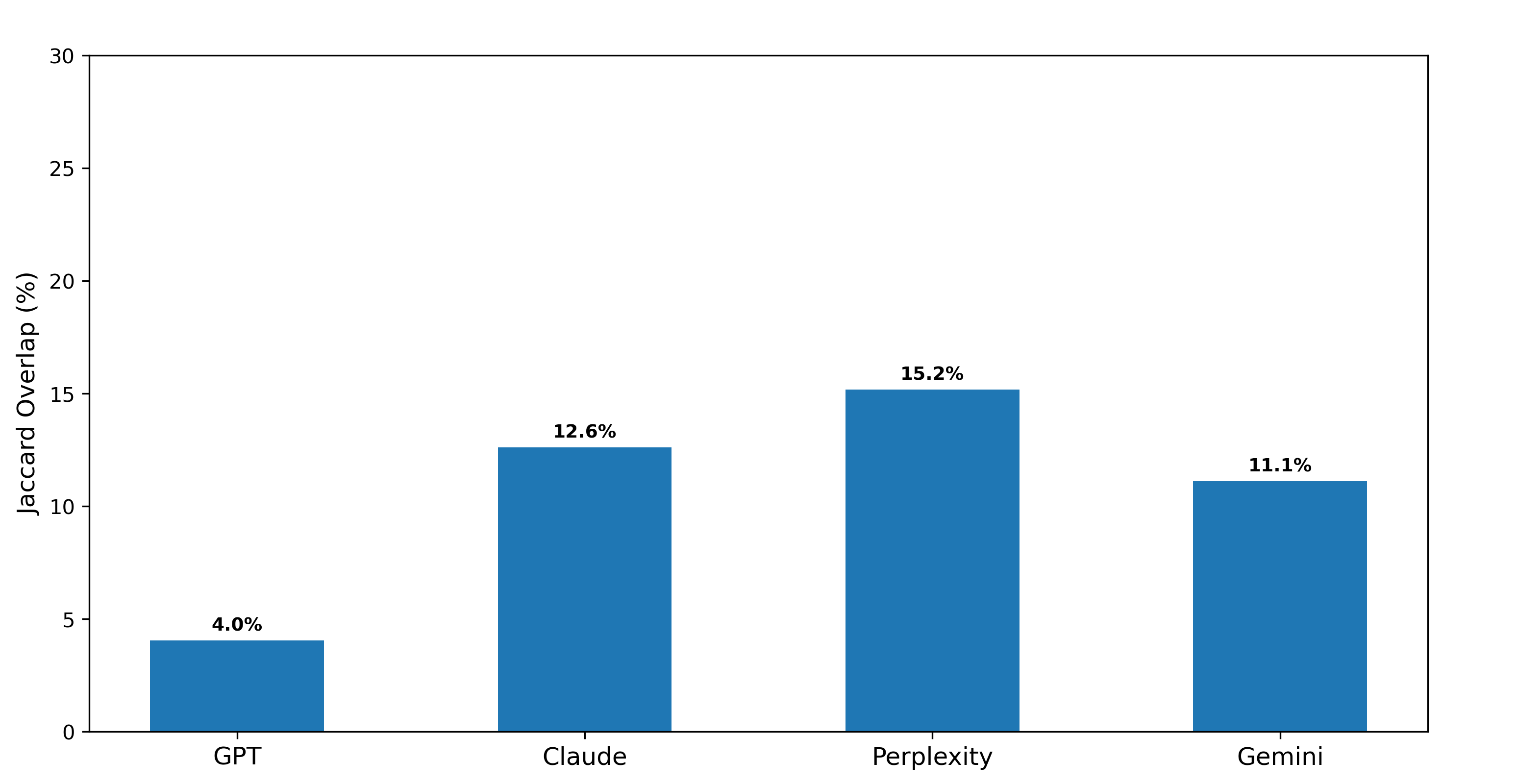}
        \caption{AI-vs-Google domain overlap over ranking queries}
        \label{fig:ai-google-overlap}
    \end{subfigure}
    \hfill
    \begin{subfigure}[t]{0.49\linewidth}
        \centering
        \includegraphics[width=\linewidth,height=4.2cm,keepaspectratio]{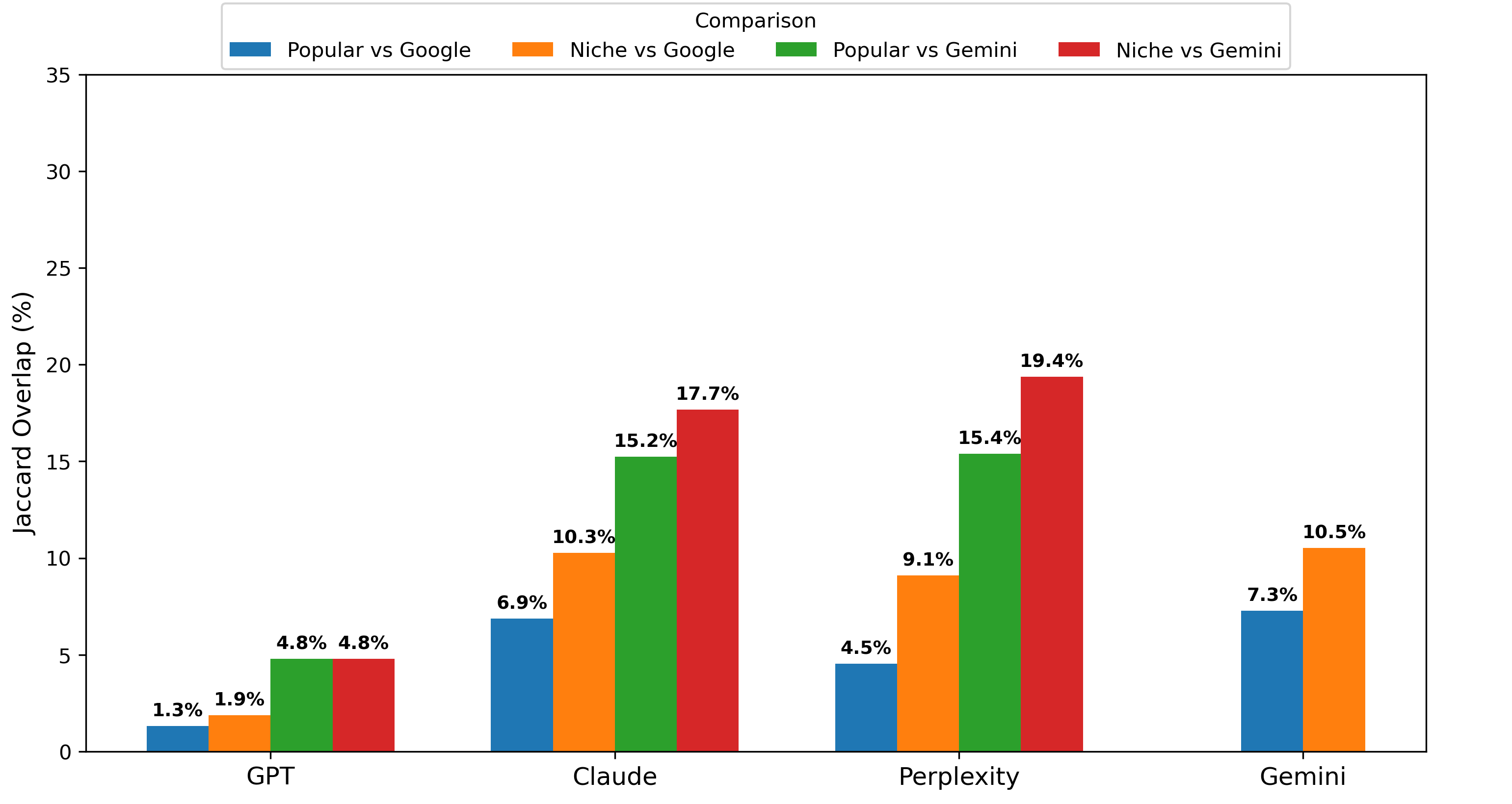}
        \caption{AI-vs-Google domain overlap on popular and niche entities}
        \label{fig:combined_overlap_by_model}
    \end{subfigure}
    \caption{Domain overlap comparisons between AI systems and search engines.}
    \label{fig:overlap-panels}
\end{figure}

Queries were generated using 100 fixed ranking-oriented templates (e.g., ``Rank the best \{topic\} from 1 to 10'', ``Experts' ranking of the best \{topic\}'', ``Best \{topic\} for most consumers''), each instantiated with the ten consumer topics, yielding 1,000 total queries. All systems were queried using identical prompts without personalization, and results were collected within the same time window to minimize temporal drift. For each query, we extract the top-10 URLs returned by Google Search and all cited URLs returned by each AI system, normalize them to their registrable domains, and compute the Jaccard overlap between each model’s domain set and Google’s top-10 results, then average the resulting values across all queries. Figure~\ref{fig:overlap-panels}(a) summarizes the cross-system overlap.

Across systems, we observe uniformly low domain-level overlap with Google's results, highlighting that generative engines surface markedly different domain ecosystems. GPT-4o shows the lowest mean overlap at 4.0\% (std = 6.6\%), followed by Gemini 2.5 Flash (11.1\%, std = 10.2\%), Claude 4.5 Sonnet (12.6\%, std = 12.2\%), and Perplexity Sonar Pro (15.2\%, std = 11.6\%). All pairwise differences in mean overlap between systems are statistically significant under paired bootstrap resampling over the same query set (10,000 iterations; all $p < 0.001$), confirming that these divergences are consistent across queries. GPT-4o’s substantially lower overlap indicates that its cited domains diverge the most from Google’s top-ranked domains. Notably, the median overlap for GPT-4o is 0.0\%, indicating that for more than half of the queries, no domain overlap with Google’s top-10 results is observed; in contrast, median overlaps for Gemini 2.5 Flash, Claude 4.5 Sonnet, and Perplexity Sonar Pro are 8.5\%, 8.7\%, and 14.3\%, respectively. Together, these results establish that AI and traditional search operate over distinct source landscapes, motivating a deeper examination of these differences across query types and source characteristics.

\paragraph{Effect of entity popularity.}
We compare 216 entity-comparison queries (108 popular, 108 niche) to examine domain overlap variations. Popular queries consist of comparisons between two globally recognized consumer brands (e.g., ``Nike or Adidas: which is better? Answer with one brand name.''), typically asking for an overall preference between the two entities. In contrast, niche queries compare two less mainstream or more specialized brands within a specific use case (e.g., ``Aeropress or Chemex: which is better for coffee? Answer with one brand name.''). All queries follow the same comparison frame (``A or B: which is better?''), with niche queries adding a task-specific qualifier when needed (e.g., ``for X''). Both popularity categories span a range of consumer domains, such as consumer electronics and home appliances.

Domain overlap is computed using the same normalization and Jaccard protocol described in the \textit{Domain overlap} analysis above. In addition to AI--Google overlap, we also compute pairwise domain overlap between each AI system and Gemini 2.5 Flash.

Figure~\ref{fig:overlap-panels}(b) presents low overall overlap, but niche queries increase alignment by 3-4 percentage points for most models. GPT-4o shows minimal increase (1.3\% to 1.9\%) but maintains the lowest overlap overall, indicating distinct sourcing patterns. This increase is statistically significant for Claude 4.5 Sonnet, Gemini 2.5 Flash, and Perplexity Sonar Pro under bootstrap resampling over queries within the two popularity groups (10,000 iterations; all $p < 0.01$), while GPT-4o exhibits a smaller and statistically non-significant increase.

The niche query overlap increase reflects narrower topical scope, concentrating sources to specific review sites and discussion threads. This is supported by declining unique-domain ratios (74.2\% to 68.6\%) and slight cross-model overlap increases (+1.1\%). AI engines thus rely on concentrated, shared sources for niche entities rather than converging toward Google's ranking logic. Similar observations hold when comparing to the results of Gemini 2.5 Flash.

\subsection{Typology of Cited Sources}

Beyond domain overlap, we analyze source types: brand (official or company-owned sites), earned (independent media and review outlets), and social (user-generated or community platforms). We categorize sources from 300 consumer-electronics queries evenly distributed across three intent categories (100 informational, 100 consideration, 100 transactional). Informational queries are knowledge-seeking (e.g., ``How do OLED TVs work?''), consideration queries reflect comparative evaluation (e.g., ``Best budget noise-canceling headphones under \$200''), and transactional queries are purchase-oriented (e.g., ``Buy Apple AirPods Pro 2 near me''). Unlike the ranking-style queries in Section 2.1, these queries were designed to reflect realistic search formulations within a single topical domain rather than following a fixed template.

\begin{figure}[t]
    \centering
    \includegraphics[width=0.92\linewidth]{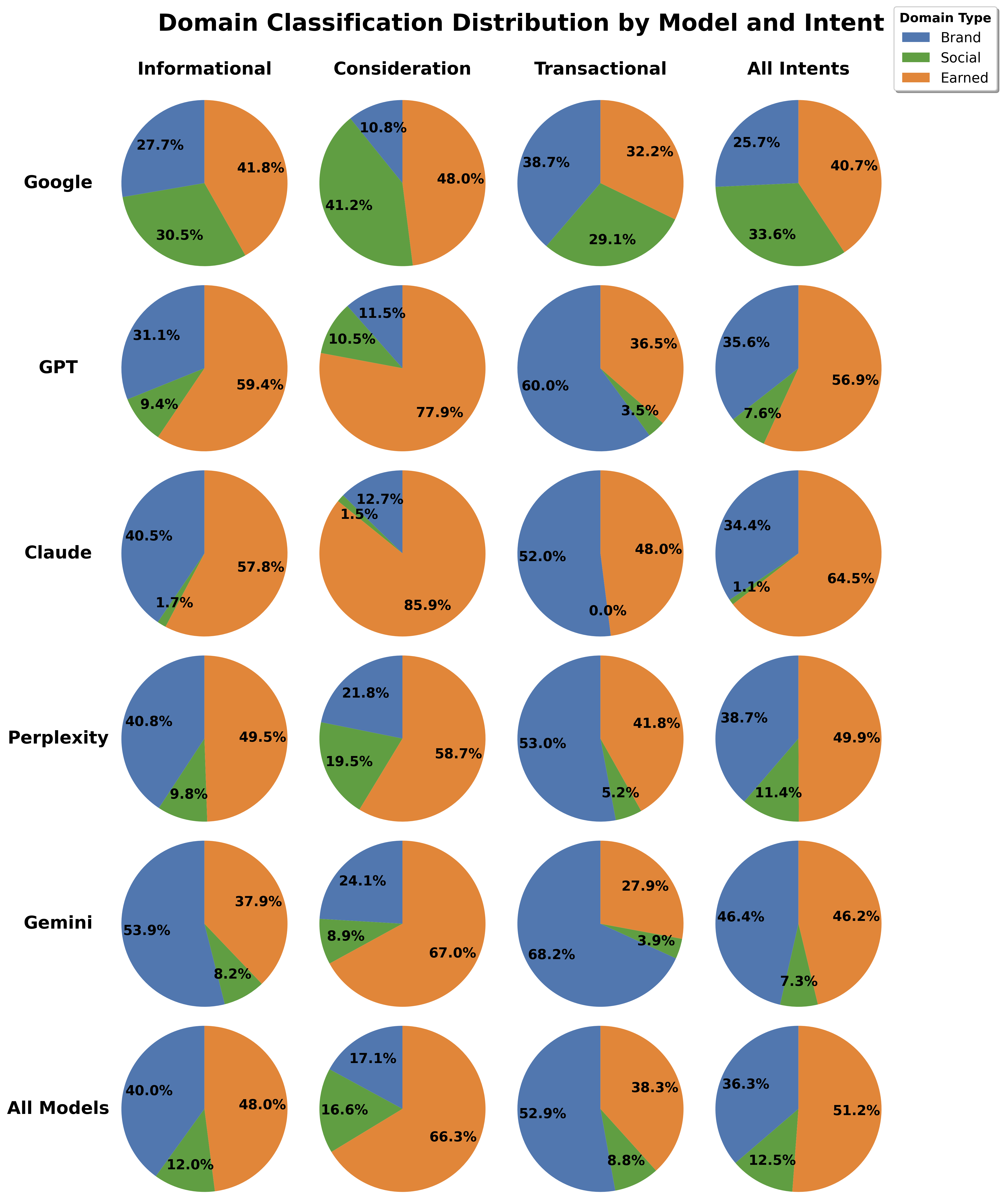}
    \caption{Source category distribution by intent and model}
    \label{fig:intent_distribution_by_model_intent_grid_all}
\end{figure}

For each cited source, we classify its source type using GPT-4o (temperature = 0) under a standardized labeling prompt restricted to the three categories: Brand, Earned, and Social. Brand refers to official company-owned domains (e.g., apple.com), Earned to independent media and review outlets (e.g., forbes.com), and Social to community or user-generated platforms (e.g., reddit.com). In addition to model-based labeling, links from predefined social media platforms are automatically assigned to the Social category to ensure consistency. A manual spot-check of a random subset confirmed high agreement with the automated labels. Figure~\ref{fig:intent_distribution_by_model_intent_grid_all} shows source type distributions by intent and system.

\paragraph{Aggregate source composition.}
Google shows balanced sourcing (41\% earned, 34\% social, 26\% brand), while AI engines favor earned over social content. Claude 4.5 Sonnet concentrates most heavily (65\% earned, 1\% social), followed by GPT-4o (57\% earned, 8\% social). Perplexity Sonar Pro (50\% earned, 39\% brand) and Gemini 2.5 Flash (46\% each earned/brand) are more balanced. Notably, Claude 4.5 Sonnet initially returned no links for most informational and transactional queries without explicit search prompting, despite being queried in web-enabled mode.

\paragraph{Variation by user intent.}
Source composition shifts systematically by intent. For informational queries, AI models vary widely in earned vs. brand emphasis, while Google maintains balance with more social content. In consideration queries, AI engines converge toward earned dominance (59-86\%), contrasting Google's social-focused approach (41\%). For transactional intent, all AI systems sharply increase brand citations (52-68\%) versus Google's mixed approach.

\paragraph{Interpretation.}
Together, these results indicate that while overlap analyses demonstrate how AI and traditional search diverge, source typology helps explain why: generative engines systematically privilege earned and brand-owned content while under-representing social and community perspectives, and their source composition varies far more sharply across intents than Google’s relatively stable profile. This intent-adaptive sourcing behavior highlights a fundamental difference in relevance logic between generative and traditional search. The next section examines whether these typological differences extend to when sources are cited, that is, their temporal freshness.

\subsection{Vertical Freshness Analysis}
\label{subsec:vertical-freshness}

We compare the temporal freshness of sources returned by three answer engines, Claude 4.5 Sonnet, GPT-4o, and Perplexity Sonar Pro, against Google across two high-interest verticals: \emph{consumer electronics} and \emph{automotive}. For each vertical, we issue a fixed set of 100 curated ranking-style queries and collect up to 10 URLs per query and engine. For Google, these are the top-$k$ ranked results returned by the search API; for answer engines, these are the URLs explicitly cited in the response.

We canonicalize URLs (strip fragments and normalize redirects when available) and deduplicate within each (engine, vertical) before computing summary statistics. We then extract a page date from HTML metadata and structured fields, including \texttt{<meta>} tags, Schema.org JSON-LD (e.g., \texttt{datePublished}, \texttt{dateModified}), \texttt{<time>} tags, and date strings in the visible body text. When multiple candidates are present, we select a single best estimate by preferring explicit publication-time signals over modification-time signals; if no usable date can be extracted, the URL is marked undated. We compute \emph{article age} as the difference (in days) between the crawl timestamp and the selected date.

We report (i) extraction coverage, defined as the fraction of collected URLs for which a date could be extracted, and (ii) the distribution of ages over dated URLs (Figure~\ref{fig:age_violin}). Because age distributions are heavy-tailed, Figure~\ref{fig:age_violin} clips ages at 365 days for readability; all reported summary statistics use the unclipped ages. To compare engines when coverage differs, we also report a coverage-adjusted freshness score
$F_{\mathrm{adj}} = F \times \mathrm{coverage}$, where
\begin{equation}
F=\frac{1}{n}\sum_{i=1}^{n}\frac{1}{1+\mathrm{age}_i}
\end{equation}
is computed over dated URLs only.

\subsubsection{Consumer electronics.}
Claude 4.5 Sonnet achieves the highest date extraction coverage (0.925; 745/805 dated URLs) and returns the freshest median content at 62.3 days. GPT-4o has comparable coverage (0.930; 623/670) with a median age of 79.8 days. Perplexity Sonar Pro dates fewer sources (0.630; 383/608) and has an older median age of 90.4 days. Google’s median age is 130.4 days with coverage 0.615 (579/941). Under the coverage-adjusted freshness score, GPT-4o ranks first, narrowly ahead of Claude 4.5 Sonnet, with Perplexity Sonar Pro third.

\subsubsection{Automotive.}
Automotive exhibits a stronger long tail of older sources across all systems. Claude 4.5 Sonnet returns the freshest median content at 148.0 days (coverage 0.609; 515/845). GPT-4o follows with a median of 162.2 days (coverage 0.734; 477/650). Perplexity Sonar Pro both dates fewer sources (0.426; 280/657) and returns older median content at 216.6 days. Google is substantially older, with a median age of 492.9 days and coverage 0.443 (413/932). Under the same coverage-adjusted freshness score, Claude 4.5 Sonnet ranks first, followed by GPT-4o and Perplexity Sonar Pro.

\subsubsection{Interpretation}
\begin{figure}[t]
\centering
\includegraphics[width=\linewidth]{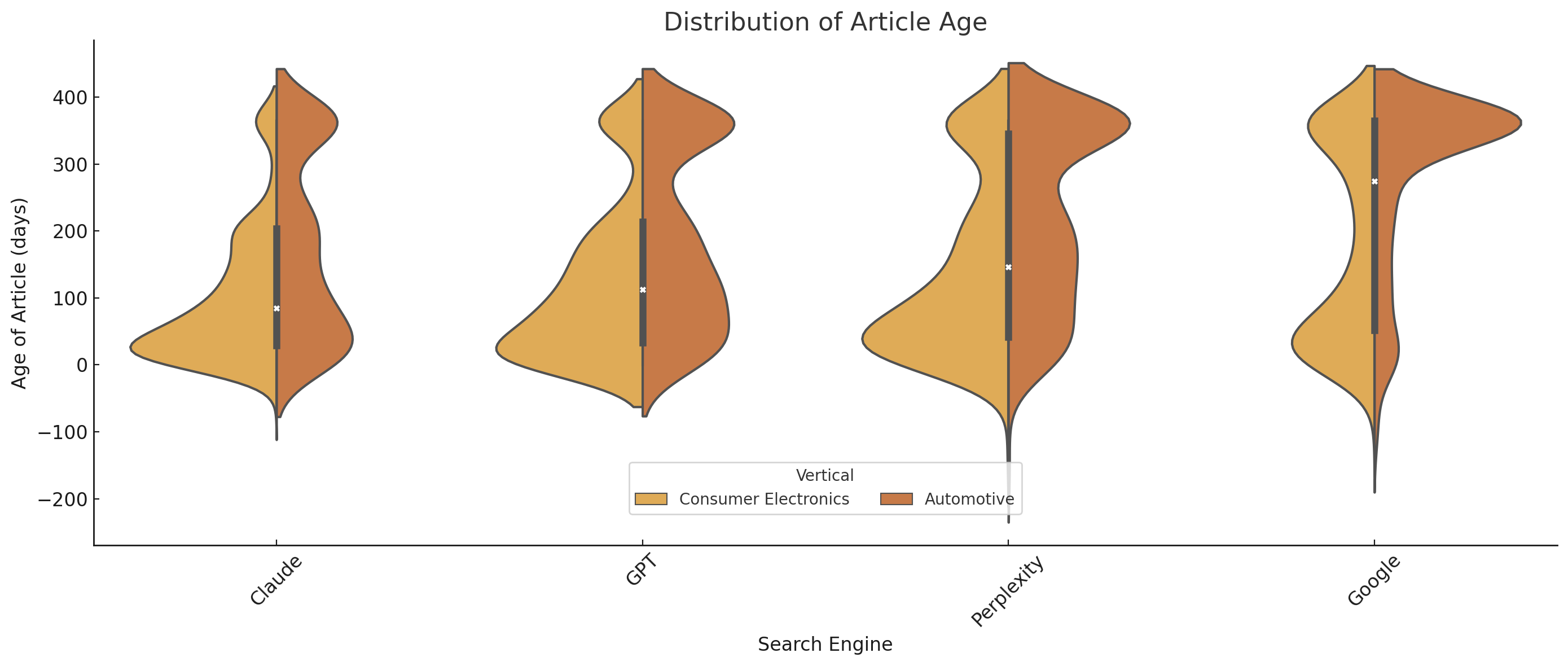}
\caption{Distribution of article age (days) by engine and vertical; ages are clipped at 365 days for readability.}
\label{fig:age_violin}
\end{figure}

\begin{figure}[t]
\centering
\begin{subfigure}[t]{0.49\linewidth}
    \centering
    \includegraphics[width=\linewidth]{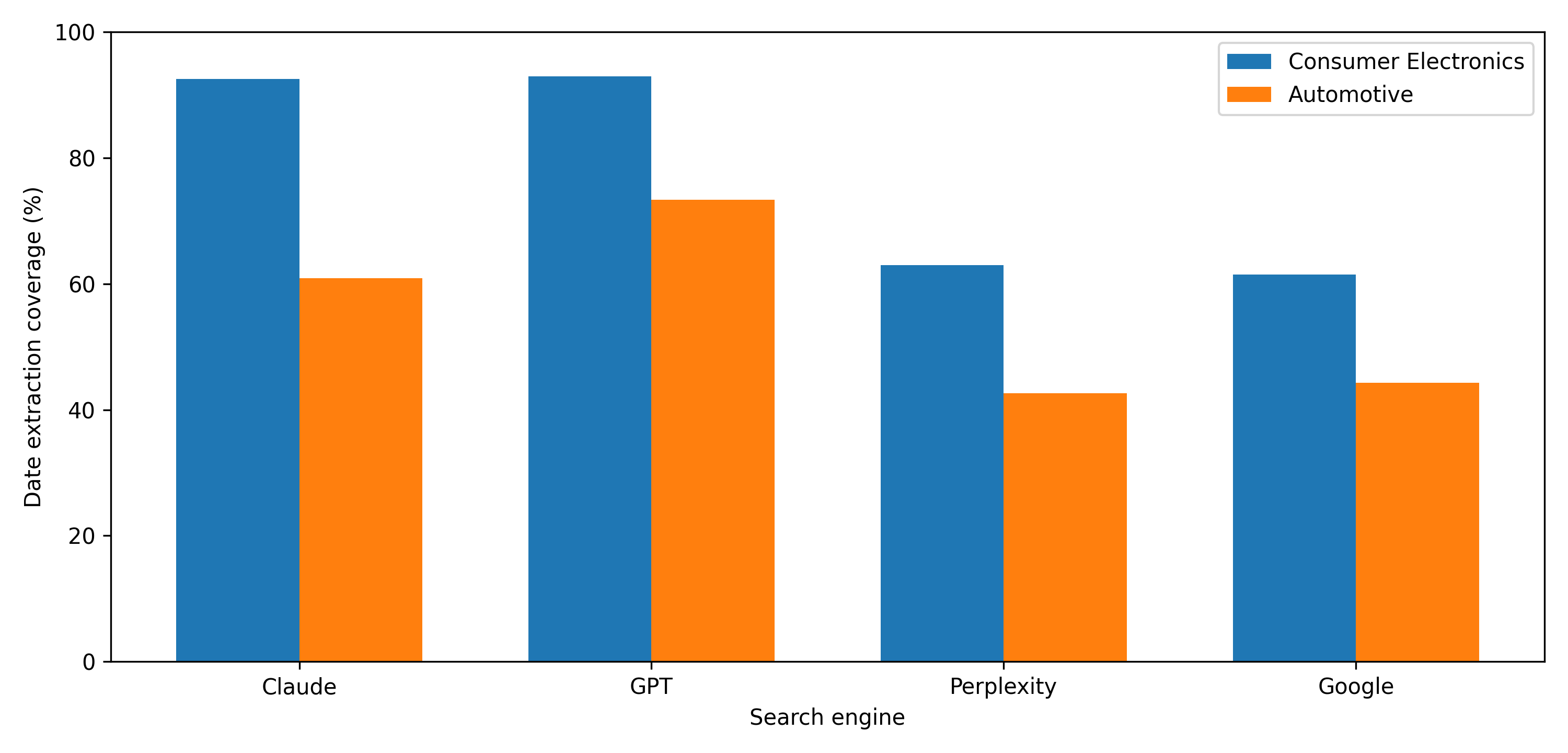}
    \caption{Date-extraction coverage (dated URLs / collected URLs) by engine and vertical}
    \label{fig:coverage}
\end{subfigure}
\hfill
\begin{subfigure}[t]{0.49\linewidth}
    \centering
    \includegraphics[width=\linewidth]{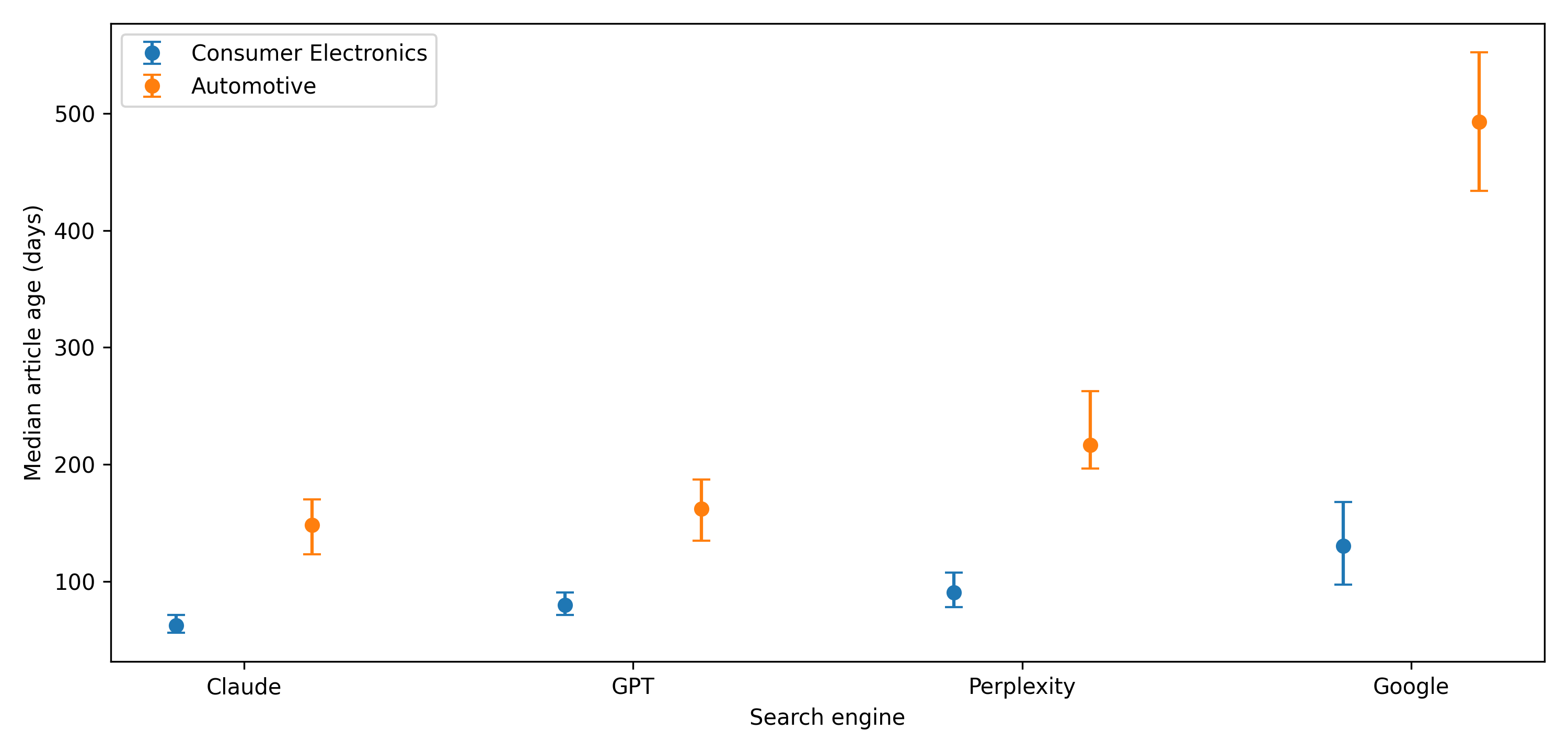}
    \caption{Median article age (days) with 95\% bootstrap confidence intervals computed over the dated URLs for each engine and vertical}
    \label{fig:median_ci}
\end{subfigure}
\caption{Freshness summary by engine and vertical.}
\label{fig:freshness-panels}
\end{figure}

Across both verticals, the answer engines return newer cited material than Google on the median, with the gap widening in automotive (148--217 days vs.\ 493 days). Coverage varies materially by engine and vertical, which matters because freshness estimates are computed over dated URLs only. We therefore report age distributions in Figure~\ref{fig:age_violin}, coverage in Figure~\ref{fig:freshness-panels}(a), and median-age summaries in Figure~\ref{fig:freshness-panels}(b), and include both raw age summaries and a coverage-adjusted freshness score when making cross-engine comparisons.

\section{The Effect of Pre-Training Bias on Answer Generation}
\label{sec:pretraining-bias}

Large language models (LLMs) often rely on a mixture of retrieved evidence and pre-existing world knowledge when generating factual responses. To disentangle these two influences, we examine how pre-training bias affects both the stability of generated rankings and the model's use of citations under different prompting and grounding conditions. Unlike Section~2, which compares several systems, this section is a single-model case study using GPT-4o so that the effects of prompting and evidence perturbations can be isolated cleanly.

\subsection{Experimental Setup}
\label{subsec:setup}

To investigate how pre-training bias interacts with retrieved evidence, we designed a controlled experimental pipeline that exposes the model to progressively perturbed inputs. We focus on ranking-oriented queries (\emph{e.g.}, ``best SUVs to buy in 2025'') because they provide an interpretable testbed for reasoning consistency across entities. All experiments were executed using the same large language model configuration (\texttt{gpt-4o}) under deterministic settings to minimize stochastic variance. The retrieved web snippets served as the evidence corpus for each query.

\subsubsection{Evidence Retrieval and Baseline Ranking.}
Given a query $q$, we first call \texttt{gpt-4o-search-preview} with web search enabled and a JSON-only prompt that returns a ranked \texttt{"list"} of candidate entities and a \texttt{"snippets"} array of verbatim excerpts with source URLs. This yields the evidence set $E_q = \{(s_j,u_j)\}_{j=1}^m$. We then pass $q$ and $E_q$ to \texttt{gpt-4o} with a ranking prompt that asks the model to produce the baseline ranked list $R$ of entities. In this baseline setting, the model has access to the retrieved snippets but is not explicitly forbidden from using prior knowledge. We refer to this condition as \emph{Normal Grounding}.

\subsubsection{Perturbation-Based Sensitivity Tests.}
To examine contextual robustness, we repeatedly modify the evidence or its presentation:
\begin{enumerate}
    \item \textbf{Snippet Shuffle (SS)} randomizes the order of snippets in $E_q$ and re-ranks. This tests whether the order of the search snippets, as influenced by generic web search ranking, affects the model's final decision.
    \item \textbf{Strict Grounding} adds an instruction that restricts reasoning to the provided snippets only and prohibits the use of prior knowledge. This aims to dampen the impact of pre-training knowledge.
    \item \textbf{Entity-Swap Injection (ESI)} randomly chooses two entities $(a,b)$ and swaps every mention of their names across all snippets before re-ranking. This tests whether the provided context directly influences the model's decision. If the model relies on the provided context, swapping entity mentions should affect the entity ranking.
\end{enumerate}

Each perturbation yields a new ranking $R_i$. We compute the mean absolute rank deviation as
\begin{equation}
\Delta_i = \frac{1}{|R|}\sum_{x\in R} \left|\mathrm{rank}_{R_i}(x) - \mathrm{rank}_{R}(x)\right|
\end{equation}
and report the average over all runs (10 runs per condition), $\Delta_{\text{avg}} = \tfrac{1}{N}\sum_i \Delta_i$.

\subsubsection{Pairwise Comparison Consistency.}
To measure the alignment between direct and decomposed reasoning, we derive an alternate ranking $R'$ through exhaustive pairwise judgments. For each entity pair $(a,b)$, the model is asked:
\emph{``Between $a$ and $b$, which is better for this query given the same documents?''}
Each entity's final score equals the number of pairwise wins. We then compute Kendall's~$\tau(R, R')$ to quantify the correlation between the holistic and pairwise-derived rankings, using $R'$ as a proxy ranking \cite{sun2024chatgptgoodsearchinvestigating, qin2024largelanguagemodelseffective}. Determining the order as a sequence of pairwise comparisons is a simpler problem for an LLM to solve and, according to prior studies \cite{ma2023zeroshotlistwisedocumentreranking, patel2024acornperformantpredicateagnosticsearch}, can more accurately reflect the true order.

\subsection{Methodology}

To measure how pre-training knowledge affects ranking behavior, we use a diverse set of ranking queries over entities. We group queries into two categories: \emph{Popular Entities} and \emph{Niche Entities}. Popular Entities are widely known, so we expect substantial exposure during pre-training (e.g., ``top 10 SUVs for a family''). Niche Entities are less likely to appear in pre-training data (e.g., ``top 10 family law firms in Toronto'').

For each query category, we run $N$ trials. In trial $i$, we sample a query $q_i$, retrieve an evidence set $E_{q_i}$, and generate a baseline ranking $R_i^{\mathrm{base}}$ using model $M$. For each perturbation type, we modify the evidence or prompting conditions to produce a perturbed ranking $\widetilde{R}_i$. We compute the trial-level sensitivity relative to the baseline and average over trials to obtain the aggregate sensitivity $\Delta_{\text{avg}}$.

To assess ranking consistency, we also derive a pairwise ranking $R'$ by summing per-entity wins from pairwise comparisons. We then compute consistency as Kendall's $\tau(R, R')$.

\subsection{Popular Entities}
\label{subsec:popular-brands}

For queries involving widely recognized entities (\emph{e.g.}, ``best SUVs to buy in 2025''), we expect the model's rankings to be largely governed by its pre-trained knowledge rather than by the retrieved evidence. We pose hundreds of such queries, conduct perturbations, and report our findings below.

\subsubsection{Robustness to Context Order}
In the snippet shuffle experiment, in which we randomly shuffle the order of retrieved snippets, the model's ranking remains highly stable despite random reordering. \emph{Normal Grounding} refers to the case where we ask the model to produce a ranking without imposing an evidence-only restriction. In this case, both pre-training knowledge and the retrieved snippets in the context window are available to the model.

\begin{table}[ht]
\centering
\small
\caption{Snippet Shuffle (SS) and ESI perturbations for popular and niche entities. $\Delta_{\text{avg}}$ is the mean absolute rank change across iterations.}
\label{tab:ss-esi}
\begin{tabular}{lccc}
\toprule
Setting & SS $\Delta_{\text{avg}}$ (Normal) & SS $\Delta_{\text{avg}}$ (Strict) & ESI $\Delta_{\text{avg}}$ \\
\midrule
Popular Entities & 2.30 & 1.52 & 2.60 \\
Niche Entities & 4.15 & 0.46 & 4.63 \\
\bottomrule
\end{tabular}
\end{table}

As shown in Table~\ref{tab:ss-esi}, popular-entity rankings are only mildly affected by snippet order or entity substitutions. The overall ranking order remains consistent, indicating that the internal entity hierarchy cultivated during pre-training is relatively invariant to lexical and contextual perturbations.

\subsubsection{Pairwise Comparison}
In the pairwise comparison experiments, the pairwise-derived ranking $R'$ was nearly identical to the one-shot direct ranking $R$ for popular entities.

\begin{table}[ht]
\centering
\small
\caption{Kendall $\tau$ between one-shot ranking $R$ and pairwise-derived ranking $R'$ under different grounding regimes.}
\label{tab:pairwise}
\begin{tabular}{lcc}
\toprule
Setting & $\tau$ (Normal) & $\tau$ (Strict) \\
\midrule
Popular Entities & 0.911 & 1.000 \\
Niche Entities & 0.556 & 0.689 \\
\bottomrule
\end{tabular}
\end{table}

For example, for the query \emph{``most reliable electric cars in 2025''}, the correlation is near perfect in the popular setting (Table~\ref{tab:pairwise}). This indicates that the model's underlying conceptual representation of entities is already stable and internally consistent, suggesting that its decision-making relies heavily on pre-trained priors. Log analysis across hundreds of related queries shows that, on average, 16\% of entities appearing in the generated rankings did not occur in any retrieved snippet. As shown in Table~\ref{tab:missrate-brief}, mainstream entities such as Toyota and Honda were almost always supported by evidence, whereas others (e.g., Cadillac, Infiniti) frequently appeared without being present in citations. This reinforces that for well-known entities, the model supplements retrieval with stored priors.

\begin{table}[ht]
\centering
\small
\caption{Representative citation-miss rates (SUV queries).}
\label{tab:missrate-brief}
\begin{tabular}{lcccccc}
\toprule
Entity & Toyota & Honda & Kia & Chevrolet & Cadillac & Infiniti \\
\midrule
Miss Rate & 0.06 & 0.03 & 0.10 & 0.26 & 0.58 & 0.73 \\
\bottomrule
\end{tabular}
\end{table}

\paragraph{Interpretation.}
For popular entities and other high-coverage domains, the model uses retrieved evidence primarily to reinforce pre-existing representations rather than to acquire new information. The retrieval context functions as confirmation, not discovery. As a result, even aggressive manipulations such as snippet reordering, entity swaps, or strict grounding constraints produce only minor deviations in the final ranking, underscoring the dominance of pre-training bias in answer generation for familiar entities.

\subsection{Niche Entities}
\label{subsec:niche-brands}

For queries involving less-established or domain-specific entities (\emph{e.g.}, $Q_{\text{law}}$: ``top lawyers in Toronto that specialize in family law''), the model's reliance on retrieval evidence becomes substantially stronger. Unlike popular categories, we expect these queries to fall outside the model's well-formed pre-training priors, leading to observable differences in reasoning behavior and ranking dynamics.

\subsubsection{Context Order Sensitivity}

In the snippet shuffle experiment, the model's ranking exhibits pronounced sensitivity to the order of evidence when applied to niche or less-popular entities, as summarized in Table~\ref{tab:ss-esi}. This large disparity demonstrates that unconstrained reasoning is highly susceptible to contextual presentation, while strict evidence-only grounding markedly stabilizes outcomes. In the same setting, ESI produces even larger movement for niche entities, confirming that for low-coverage queries the induced ranking is directly driven by the literal content of the retrieved snippets rather than by pre-training priors.

\subsubsection{Pairwise Comparison}
In the pairwise comparison experiments, the alignment between one-shot ranking $R$ and the pairwise-derived ranking $R'$ drops sharply for niche entities (Table~\ref{tab:pairwise}). For example, for a family-law query $Q_{\text{law}}$, we observe the reduced Kendall $\tau$ correlation reported in the table, reflecting a pronounced inconsistency between one-shot ranking and pairwise reasoning. Unlike the popular-entity case, the model lacks a stable internal hierarchy; its judgments fluctuate as it reassesses entity relationships on a per-comparison basis.

\paragraph{Interpretation.}
For niche or low-coverage subjects, the model enters a \emph{knowledge-seeking mode}, relying heavily on provided snippets to compensate for missing or uncertain priors. Retrieved evidence exerts a direct influence on the final ranking, and grounding constraints materially alter outcomes. This behavior suggests that retrieval is essential not merely for citation reinforcement but for the \emph{construction} of knowledge itself when pre-training coverage is sparse.

\section{Observations}

As user engagement increasingly shifts toward AI-powered search platforms, understanding the underlying ranking mechanisms of these services becomes essential. Traditional web search spawned the Search Engine Optimization (SEO) industry, dedicated to optimizing content for superior organic ranking. Our results suggest that for AI search, once a document is included within the model's context window, a factor that upstream retrieval can influence, its absolute position within that context may be less critical for certain query types. This underscores the importance of understanding principles for Answer and Generative Engine Optimization (AEO/GEO).

Content freshness emerges as a particularly important ranking factor in AI search ecosystems. Furthermore, current AI search behavior indicates that specific source types, particularly earned and owned media, contribute more strongly to search presence than others. The effects of model pre-training also prove important for certain queries, making it critical to understand when new content can materially impact different query categories. Consequently, developing analytical strategies that dissect query patterns to generate actionable content creation and placement plans will become increasingly vital for optimization success.

\section*{Declaration on Generative AI}

The author(s) did not use Generative AI tools to write or edit this paper.

\bibliography{main}

\end{document}